\pdfoutput=1
\documentclass[conference]{IEEEtran}
\IEEEoverridecommandlockouts
\usepackage[table]{xcolor} %
\usepackage[nocompress]{cite}
\usepackage{amsmath,amssymb,amsfonts}
\usepackage{algorithmic}
\usepackage{graphicx}
\usepackage{textcomp}
\usepackage[inline,shortlabels]{enumitem}
\usepackage{tcolorbox}
\usepackage{booktabs}
\usepackage{graphicx}
\usepackage{pgfplots}
\usetikzlibrary{patterns} 
\pgfplotsset{compat=1.17}
\usepackage{subcaption}
\usepackage{caption}
\usepackage[hidelinks]{hyperref}
\usepackage[capitalise]{cleveref}
\usepackage{enumerate}
\usepackage{forest}
\usepackage{url}
\usepackage{balance}
\usepackage{microtype}
\usepackage{fancyhdr}

\usetikzlibrary{shapes.geometric, arrows, positioning}

\fancypagestyle{fancycopyright}{%
    \fancyhf{}%
    \fancyhead[C]{Author copy of paper published at \emph{IEEE International Conference on Smart Computing (SMARTCOMP 2025)}\\
    \copyright2025 IEEE 
    }%
    \fancyfoot[C]{\thepage}%
}

\pgfplotsset{
    box plot/.style={
        /pgfplots/.cd,
        black,
        only marks,
        mark=-,
        mark size=\pgfkeysvalueof{/pgfplots/box plot width},
        /pgfplots/error bars/y dir=plus,
        /pgfplots/error bars/y explicit,
        /pgfplots/table/x index=\pgfkeysvalueof{/pgfplots/box plot x index},
    },
    box plot box/.style={
        /pgfplots/error bars/draw error bar/.code 2 args={%
            \draw  ##1 -- ++(\pgfkeysvalueof{/pgfplots/box plot width},0pt) |- ##2 -- ++(-\pgfkeysvalueof{/pgfplots/box plot width},0pt) |- ##1 -- cycle;
        },
        /pgfplots/table/.cd,
        y index=\pgfkeysvalueof{/pgfplots/box plot box top index},
        y error expr={
            \thisrowno{\pgfkeysvalueof{/pgfplots/box plot box bottom index}}
            - \thisrowno{\pgfkeysvalueof{/pgfplots/box plot box top index}}
        },
        /pgfplots/box plot
    },
    box plot top whisker/.style={
        /pgfplots/error bars/draw error bar/.code 2 args={%
            \pgfkeysgetvalue{/pgfplots/error bars/error mark}%
            {\pgfplotserrorbarsmark}%
            \pgfkeysgetvalue{/pgfplots/error bars/error mark options}%
            {\pgfplotserrorbarsmarkopts}%
            \path ##1 -- ##2;
        },
        /pgfplots/table/.cd,
        y index=\pgfkeysvalueof{/pgfplots/box plot whisker top index},
        y error expr={
            \thisrowno{\pgfkeysvalueof{/pgfplots/box plot box top index}}
            - \thisrowno{\pgfkeysvalueof{/pgfplots/box plot whisker top index}}
        },
        /pgfplots/box plot
    },
    box plot bottom whisker/.style={
        /pgfplots/error bars/draw error bar/.code 2 args={%
            \pgfkeysgetvalue{/pgfplots/error bars/error mark}%
            {\pgfplotserrorbarsmark}%
            \pgfkeysgetvalue{/pgfplots/error bars/error mark options}%
            {\pgfplotserrorbarsmarkopts}%
            \path ##1 -- ##2;
        },
        /pgfplots/table/.cd,
        y index=\pgfkeysvalueof{/pgfplots/box plot whisker bottom index},
        y error expr={
            \thisrowno{\pgfkeysvalueof{/pgfplots/box plot box bottom index}}
            - \thisrowno{\pgfkeysvalueof{/pgfplots/box plot whisker bottom index}}
        },
        /pgfplots/box plot
    },
    box plot median/.style={
        /pgfplots/box plot,
        /pgfplots/table/y index=\pgfkeysvalueof{/pgfplots/box plot median index}
    },
    box plot width/.initial=1em,
    box plot x index/.initial=0,
    box plot median index/.initial=1,
    box plot box top index/.initial=2,
    box plot box bottom index/.initial=3,
    box plot whisker top index/.initial=4,
    box plot whisker bottom index/.initial=5,
}

\newcommand{\boxplot}[2][]{
    \addplot [box plot median,#1] table {#2};
    \addplot [forget plot, box plot box,#1] table {#2};
    \addplot [forget plot, box plot top whisker,#1] table {#2};
    \addplot [forget plot, box plot bottom whisker,#1] table {#2};
}

\begin{document}

\title{Cosmos: A Cost Model for Serverless Workflows in the 3D Compute Continuum}

\author{
        \IEEEauthorblockN{Cynthia Marcelino}
        \IEEEauthorblockA{\textit{Distributed Systems Group} \\TU Wien, Vienna, Austria \\
            c.marcelino@dsg.tuwien.ac.at
        }
        \and
        \IEEEauthorblockN{Sebastian Gollhofer-Berger}
        \IEEEauthorblockA{\textit{Distributed Systems Group} \\TU Wien, Vienna, Austria \\
            e12024002@student.tuwien.ac.at}
        \and
        \IEEEauthorblockN{Thomas Pusztai}
        \IEEEauthorblockA{\textit{Distributed Systems Group} \\TU Wien, Vienna, Austria \\
            t.pusztai@dsg.tuwien.ac.at
        }
        \and
        \IEEEauthorblockN{Stefan Nastic}
        \IEEEauthorblockA{\textit{Distributed Systems Group} \\TU Wien, Vienna, Austria \\
            snastic@dsg.tuwien.ac.at}
    }

\maketitle

\begin{abstract}
Due to the high scalability, infrastructure management, and pay-per-use pricing model, serverless computing has been adopted in a wide range of applications such as real-time data processing, IoT, and AI-related workflows. However, deploying serverless functions across dynamic and heterogeneous environments such as the 3D (Edge-Cloud-Space) Continuum introduces additional complexity. Each layer of the 3D Continuum shows different performance capabilities and costs according to workload characteristics. Cloud services alone often show significant differences in performance and pricing for similar functions, further complicating cost management. Additionally, serverless workflows consist of functions with diverse characteristics, requiring a granular understanding of performance and cost trade-offs across different infrastructure layers to be able to address them individually.
In this paper, we present Cosmos, a cost- and a performance-cost-tradeoff model for serverless workflows that identifies key factors that affect cost changes across different workloads and cloud providers. 
We present a case study analyzing the main drivers that influence the costs of serverless workflows. We demonstrate how to classify the costs of serverless workflows in leading cloud providers AWS and GCP. 
Our results show that for data-intensive functions, data transfer and state management costs contribute to up to 75\% of the costs in AWS and 52\% in GCP.
For compute-intensive functions such as AI inference, the cost results show that BaaS services are the largest cost driver, reaching up to 83\% in AWS and 97\% in GCP. 

\end{abstract}


\thispagestyle{fancycopyright}
\pagestyle{fancy}

\begin{IEEEkeywords}
serverless, cost, edge, space, cloud, continuum
\end{IEEEkeywords}

\newcommand{\cmark}{\ding{51}} 
\newcommand{\xmark}{\ding{55}} 

\section{Introduction}

Serverless computing offers high scalability and automatic infrastructure management with fine-grained resource utilization in a pay-per-use business model~\cite{ServerlessWhyWhenHow,state-of-serverless,Cwasi2023}.
Due to its advantages, serverless computing has been widely adopted in different applications such as real-time data processing, IoT, and AI inference~\cite{AdvancingServerlessAI2024,optimus2024,ServerlessLLM}. 
Small pieces of code are wrapped in short-lived functions managed by the platform.
Typically, serverless functions are event-driven and stateless, which means they leverage external services, called Backend-as-a-Service (BaaS), to manage state and additional features, such as request routing, AI inference, and user authentication. Although functions are billed only for the execution time, the dependence on BaaS services leads to additional costs~\cite{RisePlanetServerless2023,2024selfprovisioningInfrastructure,castro2019rise,Truffle2024}.

Recently, the deployment of thousands of Low Earth Orbit (LEO) satellites with inter-satellite links (ISL) allows extending serverless computing beyond the edge and cloud into space, forming a 3D~Compute Continuum. 
This continuum allows for dynamic and efficient execution of serverless workflows that leverage the advantages of each of its three layers:
Edge computing for low-latency processing near data sources, the cloud for scalable, high-capacity computing, and LEO satellites for in-orbit processing and low-latency communication to reduce reliance on Earth-based data transfer~\cite{HyperDrive,MillionSats}.

However, reliably predicting the costs of serverless workflows remains challenging. Cloud providers impose complex pricing models for different services,
while edge and space layers of the 3D Continuum add further complexities, such as limited resources and energy constraints~\cite{sand,SEML_Case2019}.
Furthermore, functions in a workflow may vary from compute-intensive to data-intensive tasks, each with distinct resource and performance demands~\cite{goldfish2024,sand}.
Therefore, it is essential to model the costs of each function in a workflow to allow for finding an appropriate performance-cost tradeoff.


Common approaches for serverless cost estimation include:
\begin{enumerate*} [label=(\alph*)]
    \item \textit{Predictions}~\cite{PredictingCostsServerless2020,SAAF,lin2020modeling} use models, such as ML and math models to estimate costs based on historical execution data.
    This enables the estimation and analysis of costs without executing or even deploying a workflow.
    However, these high-level predictions often fail to provide detailed cost breakdowns or to identify the main drivers of higher expenses.
    \item \textit{Simulations}~\cite{COSTA,Demystifying2024,EndGame24} enable users to explore how costs behave under different parameter configurations. They offer valuable insights into performance and expenses across various workload patterns, highlighting important trade-offs. However, existing simulation tools often lack fine-grained parameters to identify which aspects contribute to higher costs.
\end{enumerate*}


Since current cost models are not detailed enough for precise performance-cost tradeoff decisions, users often err on the side of caution and incur higher costs to ensure performance.
To address this gap, we present a classification of serverless costs, focusing on isolating and identifying the main cost drivers of workflows. The Cosmos cost model enables the building of intelligent frameworks to optimize serverless costs and maximize performance.
Our main contributions include:
\begin{itemize}
    \item \emph{Cosmos: A cost and a performance-cost tradeoff model for serverless workflows} that incorporates the heterogeneity and dynamic characteristics of the 3D~Continuum. Cosmos isolates the main cost drivers while accounting for their interdependencies, providing an understanding of how different factors impact execution and cost, e.g., resource constraints, workload characteristics, communication overhead, and dynamic pricing.
    \item \emph{A cost taxonomy that classifies the main cost drivers}, enabling their identification among invocation, compute, data transfer, state management, and BaaS. This provides insights into specific cost drivers for serverless workflows across the different layers of the 3D~Continuum.
    \item \emph{A case study on different commercial cloud and edge providers}, including AWS x86, AWS ARM, AWS Lambda@Edge, and GCP. We analyze the primary cost drivers associated with each platform. Since executing experiments in space is currently impractical, we use cloud and edge experiments to systematically evaluate each cost driver and extrapolate the insights to the 3D~Continuum.
\end{itemize}

Our experiments show that data transfer and state management costs account for 75\% of AWS costs and 52\% of GCP costs for IO-intensive workloads. On the other hand, BaaS costs are the largest in compute-intensive functions, reaching up to 83\% on AWS and 97\% on GCP.
Our performance-cost model highlights the options with the best tradeoff.

\section{Illustrative Scenario \& Research Questions}\label{sec:motivation}

\subsection{Illustrative Scenario}\label{subsec:scenario}

\begin{figure}[!t]
    \centering
    \includegraphics[width=\linewidth]{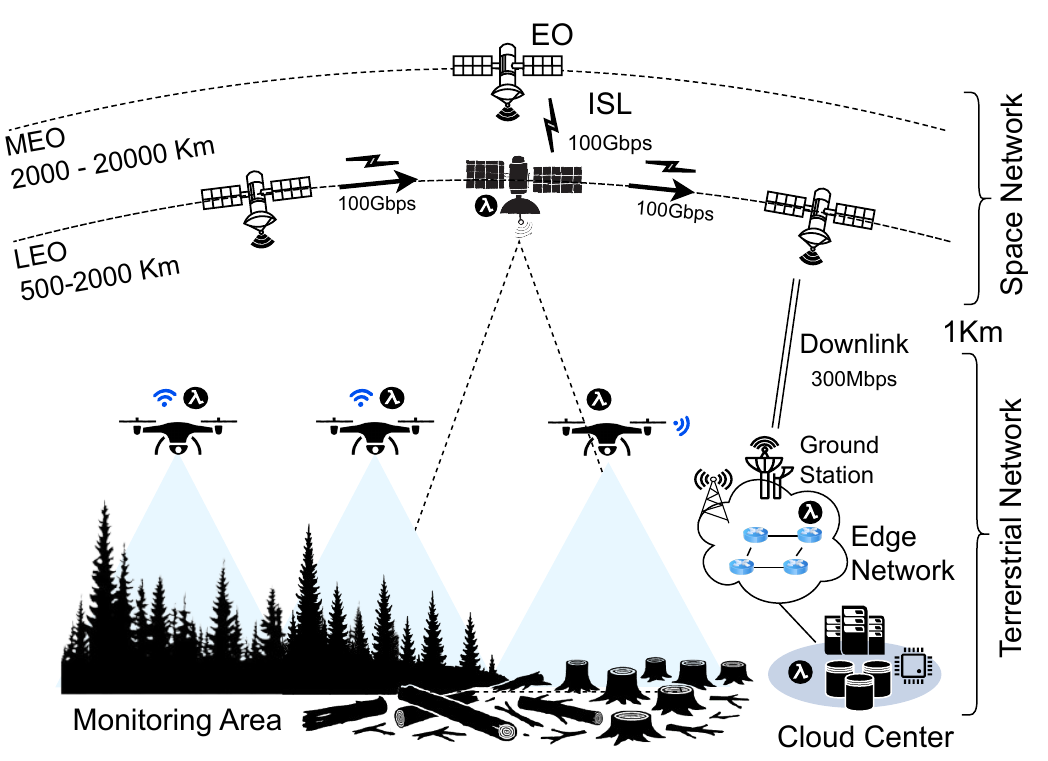}
\caption{Deforestation detection scenario with on-ground and in-orbit processing with serverless for the 3D Continuum.}
\label{fig:3d_continuum}
\end{figure}

\cref{fig:3d_continuum} shows an illustrative scenario, where the 3D~Continuum enables a scalable serverless workflow for deforestation detection in remote areas, inspired by the DETER program in Brazil~\cite{DETER2022}.
Drones collect environmental data, such as temperature, CO\textsubscript{2} levels, and high-resolution images. They transmit data to LEO satellites, which combine the edge-collected data with Earth observation~(EO) imagery in a preprocessing step directly in orbit.
Data volume reduction in space is more efficient than downlinking raw EO satellite data (1,5~TB per day~\cite{ESA_Sentinel2Ops, Sentinel2CLaunched2024}) to Earth over a radio connection with an average speed of 300~Mbps~\cite{EDRS_Overview}.
ISLs do not suffer from interference from the Earth's atmosphere and can offer bandwidths up to 100~Gbps~\cite{Network27k,DelayNoOption}. Hence, the transfer time from EO satellites is reduced.
The preprocessed data is downlinked to the cloud for performing deforestation pattern detection with compute-intensive ML models.
By properly distributing tasks across the 3D~Continuum, data can be pre-processed closer to the source, such that compute-intensive AI inference tasks in the cloud receive their inputs faster, resulting in an overall reduction in end-to-end execution time.

Optimizing data processing across all layers of the 3D~Continuum is crucial to maximizing performance without depleting resource-constrained devices such as edge devices and LEO satellites. This requires identifying the performance and cost factors within each layer to address specific limitations and enable the 3D Continuum to process the serverless workflow effectively. By understanding these factors, frameworks can dynamically allocate workloads in the most suitable locations to minimize costs while maximizing performance on resource-limited devices and reducing overall execution time.

\subsection{Research Questions}

We identify the following research questions in optimizing and classifying the costs across the 3D Continuum.

\textit{RQ-1: How can serverless workflows in the 3D~Continuum be optimized according to their workload characteristics?}

Optimizing serverless workflows across the 3D~Continuum is challenging due to the varied characteristics of each infrastructure layer.
Edge devices have limited computational power but provide low latency.
Cloud services offer high scalability, but have higher latency and dynamic pricing, which can impact cost-sensitive workflows.
LEO satellites provide low-latency network connections around the globe, but they have limited computational resources, and their power supply and onboard heat generation depend on their current position relative to the sun~\cite{HyperDrive,MillionSats,pfandzelter2021towards}.
To optimize serverless workflows, we need to identify the workload requirements and associated costs for functions, especially when workloads have different needs for computing, resource usage,  data transfer, and BaaS services.


\textit{RQ-2: How can cost models accurately capture and predict serverless execution costs for heterogeneous environments?}

Accurate prediction of serverless costs across diverse infrastructure layers requires accounting for the different characteristics, such as execution time, pricing, and operational constraints.
It is essential to identify and integrate cost drivers from the dynamic characteristics of workload and infrastructure into a unified cost model. State-of-the-art serverless cost models~\cite{Costless,Demystifying2024} do not provide fine-grained cost drivers such as fixed and dynamic prices to model the total cost of serverless functions. Failing to identify and integrate the cost drivers of dynamic environments, such as the 3D~Continuum, can lead to inaccurate cost estimates, resulting in inefficient resource allocation, reduced performance, and increased expenses.

\textit{RQ-3: How to evaluate and benchmark cost drivers for serverless functions in the heterogeneous 3D Continuum?}
Validating cost models and workloads for serverless functions requires benchmarking across different infrastructures~\cite{lin2020modeling,MLCosts2024}. However, replicating serverless workflows across edge, cloud, and space is challenging, mainly due to cross-layer interactions such as edge-to-cloud or cloud-to-space data transfers, complicating performance and cost evaluations. Therefore, isolating the impact of specific cost drivers in such heterogeneous environments is complex and expensive. 

\section{Serverless Main Cost Drivers in the 3D Continuum and Cosmos Cost and Performance-Cost Tradeoff Model}
\label{sec:model} 

\begin{figure}[t]
    \centering
\includegraphics[width=\linewidth]{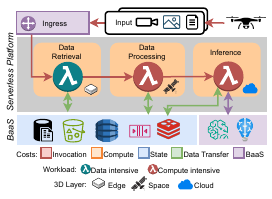}
\caption{Simplified serverless workflow for deforestation with main cost drivers along and workload characteristics of each function.}
\label{fig:faas_ml_workflow}
\end{figure}

The serverless computing pricing model allows serverless functions to be billed only for execution time, avoiding costs with idle computing resources. Typically, a serverless workflow (\cref{fig:faas_ml_workflow}) is composed of multiple functions distributed across the 3D Continuum that generate costs across several distinct factors. Each cost driver represents a specific characteristic of the task executed and services consumed during serverless workflow execution. 

\subsection{Serverless Main Cost Drivers in the 3D Continuum}

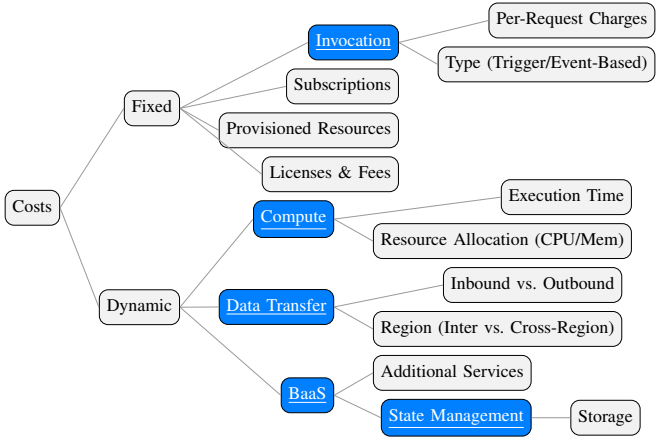
\begin{figure}[t]
    \centering
    \resizebox{\linewidth}{!}{
    \begin{forest}
    for tree={
      font=\footnotesize,
      grow'=east,
      parent anchor=east,
      child anchor=west,
      align=left,
      l sep+=2mm,
      s sep-=0.3em,
      anchor=east,
      draw, rounded corners, fill=gray!10, 
      edge={gray!80} 
    }
    [Costs
      [Fixed
        [\underline{Invocation},fill=blue!50!cyan, text=white
          [Per-Request Charges]
          [Type (Trigger/Event-Based)]
        ]
        [Subscriptions]
        [Provisioned Resources]
        [Licenses \& Fees]
      ]
      [Dynamic
        [\underline{Compute},fill=blue!50!cyan, text=white
        [Execution Time]
        [Resource Allocation (CPU/Mem)]
      ]
        [\underline{Data Transfer},fill=blue!50!cyan, text=white
              [Inbound vs. Outbound]
              [Region (Inter vs. Cross-Region)]
            ]
        [\underline{BaaS},fill=blue!50!cyan, text=white
            [Additional Services]
            [\underline{State Management}, fill=blue!50!cyan, text=white
              [Storage]
        ]
        ]
      ]
    ]
    \end{forest}
    }
 \caption{Serverless workflow costs drivers, highlighting key cost drivers: Invocation, Compute, Data Transfer, and State Management (partial view).}
\label{fig:taxonomy}
\end{figure}

\cref{fig:taxonomy} presents a taxonomy of the main cost drivers associated with serverless workflows, highlighting the focus of this analysis: Invocation, Compute, Data Transfer, and State Management. The main cost drivers are directly tied to the execution and performance of serverless functions, representing the most variable and impactful cost components in typical serverless workflows. Unlike some fixed costs, such as subscriptions and provisioned resources, which remain constant regardless of usage, the underlined drivers exhibit cost fluctuations based on function activity, data flows, and resource consumption. 

\paragraph{Invocation} Invocation is the cost incurred each time a serverless function is triggered. This cost is calculated on a per-request basis and remains constant, regardless of the size of the request payload or the execution time of the function. The number of invocations or requests can influence the serverless platform's decision on scaling the function up or down.

\paragraph{Compute} Compute costs are determined by the execution time of the function and the allocated computational resources. It includes pricing based on the execution time in seconds and the memory allocated to the function, often represented in GB-seconds. The cost is directly proportional to the intensity of computation and the duration for which resources are consumed during each invocation.

\paragraph{State Management} Serverless functions are by design stateless, which means they leverage external services to store and manage state. As shown in \cref{fig:taxonomy}, State Management is also a BaaS. However, as detailed in \cref{sec:exp}, state management constitutes a significant portion of the overall function cost. It involves the persistence and handling of required data for executing serverless functions. These costs arise from storage retention, which may be fixed (e.g., monthly storage) or dynamic (e.g., per-operation costs). Typically, serverless functions leverage many state management services, such as object storage, key-value store (KVS), message brokers, and databases.

\paragraph{Data Transfer} Data transfer refers to the data movement between serverless functions and external systems. This includes both inbound and outbound data traffic, which may involve transferring data among serverless functions, external databases, and clients. Typically, the pricing model for data transfer is based on the volume of data transferred, measured in gigabytes (GB), and the network path used, such as intra-region and inter-region transfers.

\paragraph{BaaS} BaaS costs refer to the charges associated with additional services that support serverless functions and workflows. These services include managed APIs, event gateways, data processing frameworks (such as Glue DPU), and AI platforms such as AWS SageMaker and Vertex AI. The costs for these additional services can be divided into two categories: fixed costs, associated with hourly or monthly fees for service availability, and dynamic costs, which change based on the number of requests, the amount of data processed, or specific operations performed. 

\subsection{Cosmos Cost Model}

The total cost of a serverless workflow is represented as the sum of each cost driver: invocation, computation, state management, data transfer, and BaaS~\cite{MLCosts2024, Truffle2024, sonic}. Cosmos proposes to isolate the main cost drivers of serverless workflows to better understand their impact, even though these costs are interconnected. For example, while data transfer costs can affect execution time for large inputs, our model distinguishes between these factors to determine whether variations in compute time are due to the complexity of the workload or the overhead related to data movement.

\paragraph{Function Invocation Cost}  The invocation cost $C^\text{ inv}$ for function $i$ accounts for the fixed price incurred for each request handled by the function, where $n_i$ is the number of requests handled by a function $i$, and $p_{\text{inv},i}$, the price per invocation for the function $i$. Thus, the invocation cost can be expressed as:
\begin{equation}
C_{i}^\text{ inv} = n_i \cdot p_{\text{inv},i}
\end{equation}

\paragraph{Compute Cost} 
The compute cost $C^\text{ exec}$ for function $i$ depends on the execution time and computational resources consumed, where $n_i$ is the number of requests received by function $i$, $t_i$, the compute time per request, and $p_{\text{exec},i}$, the price per GB-second for execution. The data transfer for each request is calculated in \cref{eq:data_transfer}. Thus, the compute cost for a specific compute duration can be expressed as:
\begin{equation}
C_{i}^\text{ exec} = n_i \cdot t_i \cdot p_{\text{exec},i}
\end{equation}

\paragraph{State Management Costs} The state management cost $C^\text{ state}$ for function $i$ relates to additional services to store the function state such as storage systems. Typically, state management for each function $i$ includes $d_i$, the amount of data stored (typically in GB), $p_{\text{state\_fixed},i}$, the price per GB of fixed storage costs in a certain amount of time (e.g., monthly charges). Therefore, the storage cost can be expressed as:
\begin{equation}
C_{i}^\text{ state} = d_i \cdot p_{\text{state\_fixed},i}
\end{equation}

\paragraph{Data Transfer Cost}  
The data transfer cost $C^{\text{ transfer}}$ for a function accounts for the cost of transferring data both into and out of the function. It depends on the number of requests \( n_i \), handled by function \( i \), \( r_{\text{in},i} \) and \( r_{\text{out},i} \), the total input and output data size transferred per request, respectively, and \( p_{\text{t\_in},i} \) and \( p_{\text{t\_out},i} \), the respective prices per GB for input and output data transfer. The total data transfer cost can be expressed as:  

\begin{equation}  \label{eq:data_transfer}
C_{i}^\text{ transfer}  = n_i \cdot \left( r_{\text{in},i} \cdot p_{\text{t\_in},i} + r_{\text{out},i} \cdot p_{\text{t\_out},i} \right) 
\end{equation} 

\paragraph{BaaS Costs} The BaaS cost $C^\text{baas}$ for each function $i$ includes $t_{\text{fixed},i}$, the duration of fixed-cost services; $p_{\text{fixed},i}$, the price per unit time for fixed costs; $n_i$, the number of requests handled; $r_i$, the data processed per request (in GB); and $p_{\text{dynamic},i}$, the price per unit for dynamically priced services. Therefore, the BaaS cost can be expressed as:
\begin{equation}
C_{i}^\text{ baas} = t_{\text{fixed},i} \cdot p_{\text{fixed},i} + n_i \cdot r_i \cdot p_{\text{dynamic},i}
\end{equation}

\paragraph{Total Compute Layer-Specific Cost} Different layers might introduce different pricing models. For instance, LEO-based processing introduces unique computation costs for satellites, due to their high launch costs. Therefore, the total cost $C_{i,L}$ of a workflow aggregates the invocation, compute, state, data transfer, and BaaS costs for all functions in the workflow across every layer of the 3D Continuum.  It is expressed as the sum of individual costs for all functions $i$ in $F$, executed in layer $L \in \{e,c,s\}$, where $e$ edge, $c$ cloud, and $s$ space:

\begin{equation}
\begin{aligned}
C_{i,L} =
    & C_{i,L}^\text{ inv} + C_{i,L}^\text{ exec} + C_{i,L}^\text{ state} + 
    & C_{i,L}^\text{ transfer} + C_{i,L}^\text{ baas}
\end{aligned}
\end{equation}

\subsection{Cosmos Performance-Cost Trade-off Model} 

Optimizing serverless workflow execution in the 3D~Compute Continuum requires balancing two competing objectives: minimizing costs and minimizing execution time. Processing functions closer to the data source (e.g., edge or space) reduces execution time but incurs higher costs, while cloud resources are cost-effective but introduce latency. Therefore, we define an optimization model that dynamically determines function execution while respecting a given budget and latency SLO constraints.

We define the total execution time of function $i$ on layer $L$ as $T_{i,L}$
Our goal is to minimize both total cost and total execution time, where $\alpha$ and $\beta$ are weighting factors that dynamically adjust the relative importance of cost and execution time. Our performance-cost model can be defined as:

\begin{equation}
\begin{aligned}
    \quad  \min \quad & \sum_{i \in F} \sum_{L} \Big( \alpha C_{i,L} + \beta T_{i,L} \Big) \\ 
    \text{s.t.} \quad 
    & \sum_{i \in F} \sum_{L} C_{i,L} \leq B, \quad (\text{Budget constraint}) \\
    & \sum_{i \in F} \sum_{L} T_{i,L} \leq L_{\max},  (\text{Latency SLO constraint})
\end{aligned}
\end{equation}

Instead of manually selecting $\alpha$ and $\beta$, we employ a Pareto front approach~\cite{SchedulingServerlessAlgo,ParetoFront} to dynamically balance cost and execution time. We solve two separate optimization problems to determine the best-case scenarios for cost and execution time.
We first minimize cost without considering execution time. Let the cost minimization from this be $C^*$:

\begin{equation}
C^* = \min \sum_{i \in F} \sum_{L} C_{i,L} \quad \Rightarrow \quad \alpha = \frac{1}{C^*}
\end{equation}

Next, we minimize execution time and let this minimization be $T^*$:

\begin{equation}
T^* = \min \sum_{i \in F} \sum_{L} T_{i,L} \quad \Rightarrow \quad \beta = \frac{1}{T^*}
\end{equation}

Cosmos performance-cost model trade-off ensures an optimal balance in which cost and performance are equally prioritized by dynamically adjusting weighting $\alpha$ and $\beta$ based on the best achievable budget and latency SLOs, making it well-suited for optimizing serverless workflows in 3D~Compute Continuum.

\section{Case Study Implementation}
\label{sec:impl} 

We implemented a simplified serverless workflow as described in \cref{subsec:scenario}. We identified the cost drivers using AWS (x86, ARM, and Lambda@Edge) and GCP, which offer detailed billing metrics and in-depth insights. Our case study has three serverless functions: data retrieval, data processing, and AI inference. Each function is implemented in Python with the required BaaS services to emulate real-world serverless workflows and measure their cost drivers. Our case study implementation is published as an open-source framework part of the Polaris SLO Cloud project and available on Github\footnote{\url{https://github.com/polaris-slo-cloud/cosmos}}.

\paragraph{Data Retrieval} We use AWS Lambda for compute, API Gateway for HTTP requests, and retrieve data from DynamoDB or S3. In GCP, we use Cloud Functions to interact with Firestore for queries and Cloud Storage for retrieval, responding directly to HTTP requests.

\paragraph{Data Processing} In AWS, we utilize Lambda for computing, API Gateway for request handling, and AWS Glue to execute ETL tasks, with data written to S3 or DynamoDB. GCP implements Cloud Functions for orchestration and Dataflow for ETL processing, leveraging its pay-as-you-go model for CPU, memory, and data transfer. 

\paragraph{AI Inference} We use AWS Lambda for preprocessing, API Gateway for request handling, and SageMaker Serverless for inference, while in GCP, Cloud Functions route requests to Vertex AI for model execution. Invocation, execution, storage, and data transfer costs are logged via AWS CloudWatch and GCP Cloud Monitoring to validate our cost model.

\section{Evaluation}
\label{sec:exp} 

To validate our cost model, we implement and evaluate a serverless workflow containing typical tasks in image processing as described in our illustrative scenario in \cref{fig:3d_continuum}.
We executed the implemented workflow in two major leading cloud serverless platforms, AWS Lambda~\cite{awslambda} and GCP Google Cloud Functions (GCF)~\cite{googlefunctions}.


\paragraph{Metrics} \emph{Performance-Cost Trade-off} shows the trade-offs between lower latency and lower costs. \emph{Latency} shows the mean execution time for each function from the HTTP client. \emph{Cost} evaluates the financial costs of the functions by analyzing function invocation, execution costs, and data transfers, as well as the costs associated with BaaS services. The costs are calculated in USD per million requests.

\paragraph{Baselines} In our experiments, we validate our cost model and compare the designed metrics for our serverless functions between two leading cloud providers, AWS and GCP.

\subsection{Experimental Setup}

We executed the workflow presented in ~\cref{fig:faas_ml_workflow} using services from AWS and GCP. For each function, we selected similar services, such as Cloud Storage and AWS S3. 
We deployed AWS functions with 128MB RAM in the \texttt{eu-central-1 (Frankfurt)}  region, and GCP functions in the \texttt{europe-west3-a (Frankfurt)} region. We executed HTTP requests using the Postman REST client on a MacBook Pro 2020 with an i7 processor. 
To ensure consistency, for each experiment presented, we performed the tests five times sequentially and at similar times for both providers and calculated the mean results to analyze performance and costs under varying workload scenarios.

\subsection{Performance-Cost Trade-off Results}

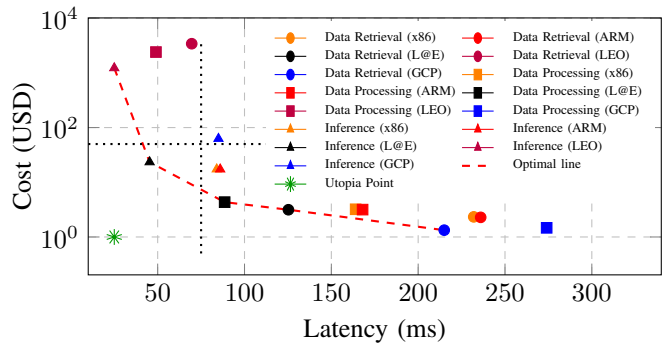
\begin{figure}
    \centering
    \begin{tikzpicture}
        \begin{axis}[
            width=9.2cm,
            height=5cm,
            ylabel style={yshift=-5pt},
            xlabel={Latency (ms)},
            ylabel={Cost (USD)},
            grid=both,
            xmax=300,
            xmin=10,
            ymin=-30,
            xmax=340,
            grid style=dashed,
            ymode=log,
            legend pos=north east,
            legend style={font=\tiny, at={(0.99,0.99)}, anchor=north east, column sep=5pt, 
                          legend image post style={xscale=0.7}, legend cell align={left}, 
                          row sep=-2pt, draw=none, /tikz/every even column/.append style={column sep=0cm}},
            legend columns=2
        ]
            \addplot[color=orange, mark=*] coordinates {(232, 2.331)};
            \addplot[color=red, mark=*] coordinates {(236, 2.2847)};
            \addplot[color=black, mark=*] coordinates {(125.28, 3.14685)};
            \addplot[color=purple, mark=*] coordinates {(69.6, 3410)};
            \addplot[color=blue, mark=*] coordinates {(215, 1.3324)};
            
            \addplot[color=orange, mark=square*] coordinates {(164, 3.211)};
            \addplot[color=red, mark=square*] coordinates {(168, 3.1647)};
            \addplot[color=black, mark=square*] coordinates {(88.56, 4.33485)};
            \addplot[color=purple, mark=square*] coordinates {(49, 2401)};
            \addplot[color=blue, mark=square*] coordinates {(274, 1.47029)};
            
            \addplot[color=orange, mark=triangle*] coordinates {(84, 17.3086)};
            \addplot[color=red, mark=triangle*] coordinates {(86, 17.2623)};
            \addplot[color=black, mark=triangle*] coordinates {(45.36, 23.36661)};
            \addplot[color=purple, mark=triangle*] coordinates {(25, 1225)};
            \addplot[color=blue, mark=triangle*] coordinates {(85, 62.5884)};
            
            \addplot[dashed, color=red, thick] coordinates {
                (25, 1225)
                (45.36, 23.36661)
                (88.56, 4.33485)
                (125.28, 3.14685)
                (215, 1.3324)
            };
            
            \addplot[color=green!60!black, mark=10-pointed star, mark size=3pt] coordinates {(25, 1.0)};
            
            \addplot[dotted, color=black, thick] coordinates {(75, 0.5) (75, 4100)};
            \addplot[dotted, color=black, thick] coordinates {(10, 50) (300, 50)};
            

            \legend{Data Retrieval (x86), Data Retrieval (ARM), Data Retrieval (L@E), Data Retrieval (LEO), Data Retrieval (GCP), 
                    Data Processing (x86), Data Processing (ARM), Data Processing (L@E), Data Processing (LEO), Data Processing (GCP), 
                    Inference (x86), Inference (ARM), Inference (L@E), Inference (LEO), Inference (GCP), Optimal line, Utopia Point}
        \end{axis}
    \end{tikzpicture}
    \caption{Cosmos Performance Cost Tradeoff Model highlighting the optimal line between latency and costs, with theoretical lowest cost and latency, and SLO constraints of 50 USD and 75ms (pointed line).}
    \label{fig:performance-cost}
\end{figure}


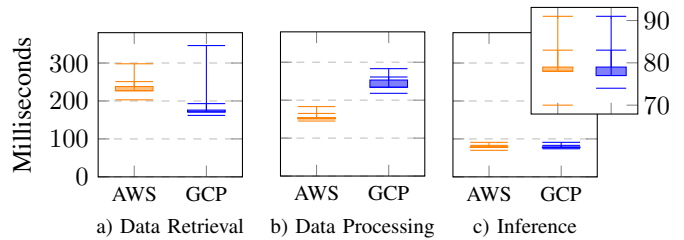
\begin{figure}[t]
\centering
\begin{tikzpicture}
        \begin{axis} [
            box plot width=2.5mm,
            width=3.5cm,
            height=3.5cm,
            ymin=0,
            ymax=380,
            ylabel=Milliseconds,
            xtick={-1, 0},
            xticklabels={AWS,GCP},
            xlabel={a) Data Retrieval},
            xlabel style={font=\footnotesize},
            ymajorgrids=true,
            grid style=dashed,
            xticklabel style={font=\footnotesize},
            enlarge x limits=0.5
        ]
        \boxplot [orange, fill=orange!50] {uc1aws.dat}
        \boxplot [color=blue, fill=blue!50] {uc1gcp.dat}
        
        \end{axis}
\end{tikzpicture}
\begin{tikzpicture}
        \begin{axis} [
            box plot width=2.5mm,
            width=3.5cm,
            height=3.5cm,
            ymin=0,
            ymax=380,
            yticklabels={},
            ylabel={},
            xtick={-1, 0},
            xticklabels={AWS,GCP},
            xlabel={b) Data Processing},
            xlabel style={font=\footnotesize},
            ymajorgrids=true,
            grid style=dashed,
            xticklabel style={font=\footnotesize},
            enlarge x limits=0.5
        ]
        \boxplot [orange, fill=orange!50] {uc2aws.dat}
        \boxplot [color=blue, fill=blue!50] {uc2gcp.dat}
        \end{axis}
\end{tikzpicture}
\hspace{-1em}
\begin{tikzpicture}
        \begin{axis} [
            box plot width=2.5mm,
            width=3.5cm,
            height=3.5cm,
            ymin=0,
            ymax=380,
            yticklabels={},
            ylabel={},
            xtick={-1, 0},
            xticklabels={AWS,GCP},
            xlabel={c) Inference},
            xlabel style={font=\footnotesize},
            ymajorgrids=true,
            grid style=dashed,
            xticklabel style={font=\footnotesize},
            enlarge x limits=0.5,
            name=mainplot
        ]
        \boxplot [orange, fill=orange!50] {uc3aws.dat}
        \boxplot [color=blue, fill=blue!50] {uc3gcp.dat}
        \end{axis}

        \node[anchor=north, inner sep=0pt] at ([yshift=0.35cm,xshift=1cm]mainplot.north) {%
                \begin{tikzpicture}
                    \begin{axis}[
                        box plot width=2mm,
                        width=3cm,
                        height=3cm,
                        xtick={-1, 0},
                        xticklabels={},
                        grid style=dashed,
                        axis background/.style={fill=white},
                        x tick label style={yshift=-0.2em},
                        y tick label style={xshift=0.2em},
                        ytick pos=right,
                        enlarge x limits=0.5
                    ]    
                \boxplot [orange, fill=orange!50] {uc3aws.dat}
                \boxplot [color=blue, fill=blue!50] {uc3gcp.dat}
        
                \end{axis}
                \end{tikzpicture}
            };
\end{tikzpicture}
\caption{End-to-end latency for serverless workflows for AWS(x86) and GCP}

\label{fig:latency}
\end{figure}

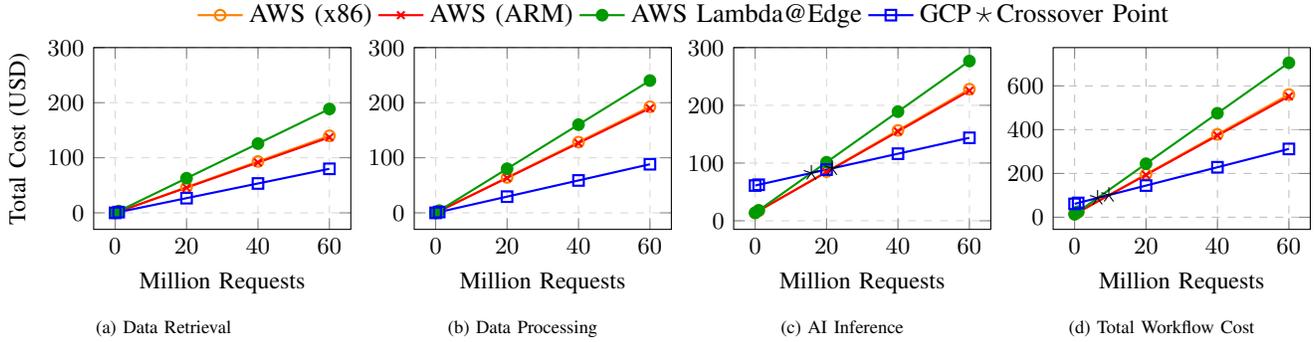
\begin{figure*}[t]
\centering
\begin{subfigure}[t]{0.24\textwidth}
\begin{tikzpicture}
\begin{axis}[
    width=5cm,
    height=4cm,
    ymax=300,
    xlabel={Million Requests},
    ylabel={Total Cost (USD)},
    xtick={0,20,40,60,80,100},
    legend style={at={(2.3,1.3)},anchor=north,legend columns=-1,draw=none},
    grid=both,
    grid style={dashed,gray!30},
    tick label style={font=\small},
    label style={font=\small},
    title style={font=\small},
    cycle list name=color list
]

\addplot[
    color=orange,
    mark=o,
    thick
]
coordinates {
    (0,0)
    (1,2.331)
    (20,46.62)
    (40,93.24)
    (60,139.86)
};
\addlegendentry{AWS (x86)}

\addplot[
    color=red,
    mark=x,
    thick
]
coordinates {
    (0,0)
    (1,2.2847)
    (20,45.694)
    (40,91.388)
    (60,137.082)
};
\addlegendentry{AWS (ARM)}

\addplot[
    color=green!60!black,
    mark=*,
    thick
]
coordinates {
    (0,0)
    (1,3.1431)
    (20,62.862)
    (40,125.724)
    (60,188.586)
};
\addlegendentry{AWS Lambda@Edge}

\addplot[
    color=blue,
    mark=square,
    thick
]
coordinates {
    (0,0)
    (1,1.3324)
    (20,26.648)
    (40,53.296)
    (60,79.944)
};
\addlegendentry{GCP}

\addlegendimage{mark=star,  mark size=3pt, color=black, only marks}
\addlegendentry{Crossover Point}
\end{axis}
\end{tikzpicture}
\caption{Data Retrieval}
\label{fig:cost_retrieval}
\end{subfigure}
\hspace{0.5em}
\begin{subfigure}[t]{0.24\textwidth}
\begin{tikzpicture}
\begin{axis}[
    width=5cm,
    height=4cm,
    ymax=300,
    xlabel={Million Requests},
    legend style={at={(0,1.3)},anchor=north,legend columns=-1,draw=none},
    grid=both,
    grid style={dashed,gray!30},
    tick label style={font=\small},
    label style={font=\small},
    title style={font=\small},
    cycle list name=color list
]

\addplot[
    color=orange,
    mark=o,
    thick
]
coordinates {
    (0,0)
    (1, 3.211 * 1)
    (20, 3.211 * 20)
    (40, 3.211 * 40)
    (60, 3.211 * 60)
};

\addplot[
    color=red,
    mark=x,
    thick
]
coordinates {
    (0,0)
    (1, 3.1647 * 1)
    (20, 3.1647 * 20)
    (40, 3.1647 * 40)
    (60, 3.1647 * 60)
    
};

\addplot[
    color=green!60!black,
    mark=*,
    thick
]
coordinates {
    (0,0)
    (1, 4.0031 * 1)
    (20, 4.0031 * 20)
    (40, 4.0031 * 40)
    (60, 4.0031 * 60)
    
};

\addplot[
    color=blue,
    mark=square,
    thick
]
coordinates {
    (0,0)
    (1, 1.47029 * 1)
    (20, 1.47029 * 20)
    (40, 1.47029 * 40)
    (60, 1.47029 * 60)
};
\end{axis}

\end{tikzpicture}
\caption{Data Processing}
\label{fig:cost_processing}
\end{subfigure}
\hspace{-1em}
\begin{subfigure}[t]{0.24\textwidth}
\begin{tikzpicture}
\begin{axis}[
    width=5cm,
    height=4cm,
    ymax=300,
    xlabel={Million Requests},
    xtick={0,20,40,60,80,100},
    legend style={at={(0.5,1.3)},anchor=north,legend columns=-1,draw=none},
    grid=both,
    grid style={dashed,gray!30},
    tick label style={font=\small},
    label style={font=\small},
    title style={font=\small},
    cycle list name=color list
]

\addplot[
    color=orange,
    mark=o,
    thick
]
coordinates {
    (0, 13.7376)
    (1, 13.7376 + 3.571 * 1)
    (20, 13.7376 + 3.571 * 20)
    (40, 13.7376 + 3.571 * 40)
    (60, 13.7376 + 3.571 * 60)
};

\addplot[
    color=red,
    mark=x,
    thick
]
coordinates {
    (0,13.7376)
    (1, 13.7376 + 3.5247 * 1)
    (20, 13.7376 + 3.5247 * 20)
    (40, 13.7376 + 3.5247 * 40)
    (60, 13.7376 + 3.5247 * 60)
    
};

\addplot[
    color=green!60!black,
    mark=*,
    thick
]
coordinates {
    (0,13.7376)
    (1, 13.7376 + 4.3831 * 1)
    (20, 13.7376 + 4.3831 * 20)
    (40, 13.7376 + 4.3831 * 40)
    (60, 13.7376 + 4.3831 * 60)
    
};

\addplot[
    color=blue,
    mark=square,
    thick
]
coordinates {
    (0, 61.056)
    (1, 61.056 + 1.378 * 1)
    (20, 61.056 + 1.378 * 20)
    (40, 61.056 + 1.378 * 40)
    (60, 61.056 + 1.378 * 60)
};

\addplot[
    color=black,
    mark=star,
    only marks,
    mark size=3pt
]
coordinates {
    (15.7460, 82.7540)
    (21.5770, 90.7891)
};


\end{axis}

\end{tikzpicture}
\caption{AI Inference}
\label{fig:cost_inference}
\end{subfigure}
\hspace{-1em}
\begin{subfigure}[t]{0.24\textwidth}
\begin{tikzpicture}
\begin{axis}[
    width=5cm,
    height=4cm,
    xlabel={Million Requests},
    xtick={0,20,40,60,80,100},
    legend style={at={(0.5,1.3)},anchor=north,legend columns=-1,draw=none},
    grid=both,
    grid style=dashed,
    tick label style={font=\small},
    label style={font=\small},
    title style={font=\small},
    cycle list name=color list
]

\addplot[
    color=orange,
    mark=o,
    thick
]
coordinates {
    (0, 13.7376)
    (1, 22.8506)
    (20, 195.9976)
    (40, 378.2576)
    (60, 560.5176)
};

\addplot[
    color=red,
    mark=x,
    thick
]
coordinates {
   (0, 13.7376)
    (1, 22.7117)
    (20, 193.2196)
    (40, 372.7016)
    (60, 552.1836)
    
};

\addplot[
    color=green!60!black,
    mark=*,
    thick
]
coordinates {
    (0, 13.7376)
    (1, 25.2669)
    (20, 244.3236)
    (40, 474.9096)
    (60, 705.4956)
    
};

\addplot[
    color=blue,
    mark=square,
    thick
]
coordinates {
    (0, 61.056)
    (1, 65.23669)
    (20, 144.6698)
    (40, 228.2836)
    (60, 311.8974)
    
};

\addplot[
    color=black,
    mark=star,
    only marks,
    mark size=3pt
]
coordinates {
    (6.4391, 87.9759)
    (9.5936, 101.1637)
};


\end{axis}
\end{tikzpicture}
\caption{Total Workflow Cost}
\label{fig:cost_cumulative}
\end{subfigure}
\caption{Cost comparison of serverless workflows across AWS(x86, ARM, and Lambda@Edge) and GCP for different workloads, showing the total costs as a function of request volume for each function and the total workflow. }
\label{fig:cost_comparison_use_case_3}
\end{figure*}

\begin{table}[t]
\centering
\caption{Detailed cost comparison across serverless functions for AWS (x86, ARM, Lambda@Edge (L@E)) and GCP in USD for 1M requests.}
\label{tab:detailed_workflow_costs}
\resizebox{\columnwidth}{!}{%
\begin{tabular}{lp{3.5cm}lllll}
\toprule
\textbf{Workflow}      & \textbf{Component}                      & \textbf{AWS (x86)} & \textbf{AWS (ARM)} & \textbf{L@E} & \textbf{GCP} & \textbf{Cost Unit} \\ \midrule
\rowcolor[HTML]{EFEFEF} \textbf{Data} 
                       & Function Invocation                     & 0.20               & 0.20                     & 0.60                        & 0.40               & 1M requests  \\
    \textbf{Retrieval} & Function Execution                      & 0.213              & 0.1667                   & 0.6251                      & 0.2304             & GB-second  \\
\rowcolor[HTML]{EFEFEF} 
                       & API Gateway                             & 1.06               & 1.06                     & 1.06                        & -                  & 1M requests  \\
                       & DynamoDB Reads                          & 0.1345             & 0.1345                   & 0.1345                      & -                  & 1M requests  \\
\rowcolor[HTML]{EFEFEF} 
                       & DynamoDB Storage                        & 0.269              & 0.269                    & 0.269                       & -                  & GB-month     \\
                       & Firestore Reads                         & -                  & -                        & -                           & 0.046              & 1M requests  \\
\rowcolor[HTML]{EFEFEF} 
                       & Firestore Storage                       & -                  & -                        & -                           & 0.231              & GB-month     \\
                       & \textbf{Total DynamoDB/Firestore}       & 0.4035             & 0.4035                   & 0.4035                      & 0.277              & Read/Store  \\
\rowcolor[HTML]{EFEFEF} 
                       & AWS S3 Retrieval                        & 0.43                & 0.43                      & 0.43                         & -                  & 1M requests  \\
                       & AWS S3 Storage                          & 0.0245              & 0.0245                    & 0.0245                       & -                  & GB-month     \\
\rowcolor[HTML]{EFEFEF} 
                       & Cloud Storage Retrieval                 & -                  & -                        & -                           & 0.4                & 1M requests  \\
                       & Cloud Storage                           & -                  & -                        & -                           & 0.025              & GB-month     \\
\rowcolor[HTML]{EFEFEF} 
                       & \textbf{Total AWS S3/Cloud Storage}     & 0.4545              & 0.4545                    & 0.4545                       & 0.425              & Read/Store \\
                       & \textbf{Total}                          & \textbf{2.331}    & \textbf{2.2847}          & \textbf{3.1431}             & \textbf{1.3324}    & -                \\ \midrule
\textbf{Data} 
                       & Function Invocation                     & 0.20               & 0.20                     & 0.60                        & 0.40               & 1M requests  \\
\rowcolor[HTML]{EFEFEF} \textbf{Processing} 
                       & Function Execution                      & 0.213              & 0.1667                   & 0.6251                      & 0.276              & GB-second  \\
                       & API Gateway                             & 1.06               & 1.06                     & 1.06                        & -                  & 1M requests  \\
\rowcolor[HTML]{EFEFEF} & DynamoDB Reads                          & 0.1345             & 0.1345                   & 0.1345                      & -                  & 1M requests  \\

                       & DynamoDB Storage                        & 0.269              & 0.269                    & 0.269                       & -                  & GB-month     \\
\rowcolor[HTML]{EFEFEF} & Firestore Reads                         & -                  & -                        & -                           & 0.046              & 1M requests  \\

                        & Firestore Storage                       & -                  & -                        & -                           & 0.231              & GB-month     \\
\rowcolor[HTML]{EFEFEF} & \textbf{Total DynamoDB/Firestore}       & 0.4035             & 0.4035                   & 0.4035                      & 0.277              & Read/Store  \\
                        & AWS S3 Retrieval                        & 0.43                & 0.43                      & 0.43                         & -                  & 1M requests  \\
\rowcolor[HTML]{EFEFEF}  & AWS S3 Storage                          & 0.0245              & 0.0245                    & 0.0245                       & -                  & GB-month     \\
 
                       & Cloud Storage Retrieval                 & -                  & -                        & -                           & 0.4                & 1M requests  \\
\rowcolor[HTML]{EFEFEF} & Cloud Storage                           & -                  & -                        & -                           & 0.025              & GB-month     \\
 
                       & \textbf{Total AWS S3/Cloud Storage}     & 0.4545              & 0.4545                    & 0.4545                       & 0.425              & Read/Store \\
\rowcolor[HTML]{EFEFEF}  & Glue DPU (2-hour ETL)                   & 0.88               & 0.88                     & 0.88                        & -                  & Processing \\
                        & Dataflow CPU                            & -                  & -                        & -                           & 0.07325            & CPU-hour     \\
\rowcolor[HTML]{EFEFEF} & Dataflow Memory                         & -                  & -                        & -                           & 0.00465            & GB-hour      \\
                        & Dataflow Processed Data                 & -                  & -                        & -                           & 0.01439            & GB processed \\
\rowcolor[HTML]{EFEFEF} & \textbf{Total ETL Costs}                & 0.88               & 0.88                     & 0.88                        & 0.09229            & DPU/Dataflow \\ 
 & \textbf{Total}                          & \textbf{3.211}     & \textbf{3.1647}          & \textbf{4.0031}             & \textbf{1.47029}   & -                \\ \midrule
\rowcolor[HTML]{EFEFEF} \textbf{AI} 
                       & Function Invocation                     & 0.20               & 0.20                     & 0.60                        & 0.40               & 1M requests  \\
    \textbf{Inference}    & Function Execution                      & 0.213              & 0.1667                   & 0.6251                      & 0.2304             & GB-second  \\
\rowcolor[HTML]{EFEFEF}  
                       & API Gateway                             & 1.06               & 1.06                     & 1.06                        & -                  & 1M requests  \\
                       & DynamoDB Reads                          & 0.1345             & 0.1345                   & 0.1345                      & -                  & 1M requests  \\
\rowcolor[HTML]{EFEFEF} 
                       & DynamoDB Storage                        & 0.269              & 0.269                    & 0.269                       & -                  & GB-month     \\
                       & Firestore Reads                         & -                  & -                        & -                           & 0.046              & 1M requests  \\
\rowcolor[HTML]{EFEFEF} 
                       & Firestore Storage                       & -                  & -                        & -                           & 0.231              & GB-month     \\
                       & \textbf{Total DynamoDB/Firestore}       & 0.4035             & 0.4035                   & 0.4035                      & 0.277              & Read/Store  \\
\rowcolor[HTML]{EFEFEF} 
                       & AWS S3 Retrieval                        & 0.43                & 0.43                      & 0.43                         & -                  & 1M requests  \\
                       & AWS S3 Storage                          & 0.0245              & 0.0245                    & 0.0245                       & -                  & GB-month     \\
\rowcolor[HTML]{EFEFEF} 
                       & Cloud Storage Retrieval                 & -                  & -                        & -                           & 0.4                & 1M requests  \\
                       & Cloud Storage                           & -                  & -                        & -                           & 0.025              & GB-month     \\
\rowcolor[HTML]{EFEFEF} 
                       & \textbf{Total AWS S3/Cloud Storage}     & 0.4545              & 0.4545                    & 0.4545                       & 0.425              & Read/Store \\
                      
                       & SageMaker Provisioning                  & 13.7376            & 13.7376                  & 13.7376                      & -                  & Fixed monthly    \\
\rowcolor[HTML]{EFEFEF} 
                       & Vertex AI Provisioning                  & -                  & -                        & -                           & 61.056             & Fixed monthly    \\
                       & SageMaker Inference                     & 1.24               & 1.24                     & 1.24                        & -                  & request      \\
\rowcolor[HTML]{EFEFEF} 
                       & Vertex AI Inference                     & -                  & -                        & -                           & 0.20               & request      \\
                       & \textbf{Total Fixed + Variable Costs}   & \textbf{17.3086}   & \textbf{17.2623}         & \textbf{18.7807}            & \textbf{62.5884}   & -                \\ \bottomrule
\end{tabular}%
}
\end{table}

\cref{fig:performance-cost} depicts the trade-off between latency and cost for different layers and platforms in the 3D continuum, including AWS (x86, ARM, Lambda@Edge), GCP, and a hypothetical LEO. To the best of our knowledge, there is no existing pay-per-use pricing model for LEO computing yet; therefore, we assume that LEO execution price of 49USD for 1M requests per ms executed based on~\cite{tm2space_orbitlab}, while achieving lower latency than L@E~\cite{Network27k,DelayNoOption}, and L@E offers nearly 2x lower latency over AWS Lambda~\cite{Bloom2018}. Further, to better understand the trade-offs and capture layer-specific impact, we assume that functions run entirely on a single layer, each offering a similar resource capacity.
In \cref{fig:performance-cost}, the dashed red line represents the optimal cost-latency trade-off. The points below are infeasible, as no deployment can achieve both lower cost and latency, illustrated by the utopia point, which represents the lowest cost and latency but is unreachable. The points above the optimal line are feasible but always involve trade-offs of either higher cost or latency. Inference (x86, ARM, and GCP) functions do not appear on the optimal line due to their high costs, which are driven by BaaS costs, making them less cost-effective compared to Data Retrieval and Data Processing. Although LEO offers the lowest latency, it significantly increases costs. Inference (GCP), Data Processing (LEO), and Data Retrieval (LEO) significantly exceed the cost and latency SLOs, making it the least among the options evaluated.
Latency and cost results are further discussed in \cref{subsec:latency} and \cref{subsec:cost}, respectively.

\begin{tcolorbox}[colframe=gray, colback=gray!10, coltitle=black, boxrule=0.3mm,  rounded corners,left=1mm, right=1mm, top=1mm, bottom=1mm]
\textbf{\emph{Performance-Cost Takeaway:}} The optimal line in \cref{fig:performance-cost} highlights the best trade-offs, with any function above it requiring sacrifices in either cost or latency.
\end{tcolorbox}

\subsection{Latency Results}\label{subsec:latency}

\cref{fig:latency} presents the end-to-end latency results for serverless functions across three workflows: data retrieval (\cref{fig:latency}a), data processing (\cref{fig:latency}b), and inference (\cref{fig:latency}c). The $x$ axis represents the cloud providers (AWS and GCP), and the $y$ axis indicates response latency in milliseconds. For data retrieval, AWS exhibits latencies ranging from 203 ms to 298 ms, while GCP shows a range from 162 ms to 346 ms. In data processing, AWS maintains a latency range of 145 ms to 183 ms, compared to GCP’s range of 218 ms to 283 ms. For inference, AWS demonstrates a latency range of 70 ms to 92 ms, while GCP records latencies between 74 ms and 91 ms.

\begin{tcolorbox}[colframe=gray, colback=gray!10, coltitle=black, boxrule=0.3mm,  rounded corners,left=1mm, right=1mm, top=1mm, bottom=1mm]
\textbf{\emph{Latency Takeaway:}} GCP shows 13\% lower latency for data-intensive functions, while AWS outperforms GCP by 36\% in compute-intensive tasks. For AI inference, both perform similarly, with AWS slightly faster.
\end{tcolorbox}

\subsection{Cost Results}\label{subsec:cost}

\begin{figure}[t]
    \centering
    \hspace{-2em}
    \begin{subfigure}{0.3\linewidth}
        \centering
        \begin{tikzpicture}
            \begin{axis}[
                ybar stacked,
                bar width=12pt,
                width=4cm,
                height=4cm,
                ylabel={Cost (USD)},
                ylabel style={yshift=-5pt, font=\small},
                symbolic x coords={x86,ARM,L@E, GCP},
                xticklabel style={
                    font=\scriptsize
                },
                xtick=data,
                ymin=0,
                xlabel=(a),
                xlabel style={font=\scriptsize},
                xticklabel style={rotate=45},
                legend style={at={(1.65,1.3)},anchor=north,draw=none,legend columns=-1,font=\footnotesize},
                ymajorgrids=true,
                grid style=dashed,
                enlarge x limits=0.2,
                ymax=4.5
                ]
                \addplot[fill=blue!50,postaction={pattern=north west lines}] coordinates {(x86, 0.20) (ARM, 0.20) (L@E, 0.60)  (GCP, 0.40)}; 
                \addplot[fill=orange!50,postaction={pattern=north east lines}] coordinates {(x86, 0.213) (ARM,0.1667) (L@E,0.6251) (GCP, 0.2304)}; 
                \addplot[fill=green!50,postaction={pattern=crosshatch}] coordinates {(x86, 1.06) (ARM,1.06) (L@E,1.06) (GCP, 0)}; 
                \addplot[fill=red!50,postaction={pattern=grid}] coordinates {(x86, 0.5645) (ARM,0.5645) (L@E,0.5645)(GCP, 0.446)}; 
                \addplot[fill=yellow!50,postaction={pattern=crosshatch dots}] coordinates {(x86, 0.2935) (ARM,0.2935) (L@E,0.2935)(GCP, 0.256)}; 
                \legend{Invocation, Execution, BaaS, Data Transfer, State Management}
            \end{axis}
        \end{tikzpicture}
        \label{fig:breakdown_a}
    \end{subfigure}
    \hspace{1.5em}
    \begin{subfigure}{0.3\linewidth}
        \centering
        \begin{tikzpicture}
            \begin{axis}[
                ybar stacked,
                bar width=12pt,
                width=4cm,
                height=4cm,
                yticklabels={},
                symbolic x coords={x86,ARM,L@E, GCP},
                xticklabel style={
                    font=\scriptsize
                },
                xtick=data,
                ymin=0,
                xlabel=(b),
                xlabel style={font=\scriptsize},
                xticklabel style={rotate=45},
                legend style={at={(1.5,1.3)},anchor=north,draw=none,legend columns=-1,font=\footnotesize},
                ymajorgrids=true,
                grid style=dashed,
                enlarge x limits=0.2,
                ymax=4.5
                ]
                \addplot[fill=blue!50,postaction={pattern=north west lines}] coordinates {(x86, 0.20) (ARM,0.20) (L@E,0.60) (GCP, 0.40)}; 
                \addplot[fill=orange!50,postaction={pattern=north east lines}] coordinates {(x86, 0.213) (ARM, 0.1667) (L@E,0.6251) (GCP, 0.276)}; 
                \addplot[fill=green!50,postaction={pattern=crosshatch}] coordinates {(x86, 1.94) (ARM, 1.94) (L@E, 1.94) (GCP, 0)}; 
                \addplot[fill=red!50,postaction={pattern=grid}] coordinates {(x86, 0.5645) (ARM,0.5645) (L@E,0.5645)(GCP, 0.446)}; 
                \addplot[fill=yellow!50,postaction={pattern=crosshatch dots}] coordinates {(x86, 0.2935) (ARM,0.2935) (L@E,0.2935)(GCP, 0.256)}; 
            \end{axis}
        \end{tikzpicture}
        \label{fig:breakdown_b}
    \end{subfigure}%
    \hspace{-0.5em}
    \begin{subfigure}{0.3\linewidth}
        \centering
        \begin{tikzpicture}
            \begin{axis}[
                ybar stacked,
                bar width=12pt,
                width=4cm,
                height=4cm,
                symbolic x coords={x86,ARM,L@E, GCP},
                xticklabel style={
                    font=\scriptsize
                },
                xtick=data,
                ymin=0,
                xlabel=(c),
                xlabel style={font=\scriptsize},
                xticklabel style={rotate=45},
                legend style={at={(1.5,1.3)},anchor=north,draw=none,legend columns=-1,font=\footnotesize},
                ymajorgrids=true,
                grid style=dashed,
                enlarge x limits=0.2
                ]
                \addplot[fill=blue!50,postaction={pattern=north west lines}] coordinates {(x86, 0.20) (ARM,0.20) (L@E,0.60) (GCP, 0.40)}; 
                \addplot[fill=orange!50,postaction={pattern=north east lines}] coordinates {(x86, 0.213) (ARM, 0.1667) (L@E,0.6251) (GCP, 0.276)}; 
                \addplot[fill=green!50,postaction={pattern=crosshatch}] coordinates {(x86, 13.7376) (ARM, 13.7376) (L@E, 13.7376)  (GCP, 61.056)}; 
                \addplot[fill=red!50,postaction={pattern=grid}] coordinates {(x86, 0.5645) (ARM,0.5645) (L@E,0.5645)(GCP, 0.446)}; 
                \addplot[fill=yellow!50,postaction={pattern=crosshatch dots}] coordinates {(x86, 0.2935) (ARM,0.2935) (L@E,0.2935)(GCP, 0.256)}; 
            \end{axis}
        \end{tikzpicture}
        \label{fig:breakdown_c}
    \end{subfigure}
    \captionsetup{skip=-8pt}
    \caption{Cost breakdowns for AWS (x86, ARM), AWS Lambda@Edge (L@E) and GCP across different functions in a serverless workflow for 1M requests. (a) Data retrieval, (b) Data processing and (c) AI inference.}
    \label{fig:cost_breakdown}
\end{figure}
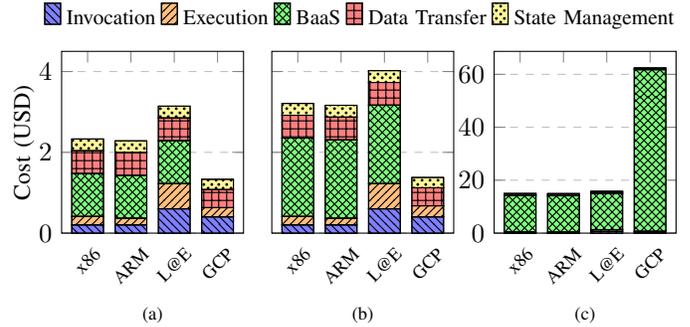

\cref{fig:cost_comparison_use_case_3} compares serverless workflow costs across AWS (x86, ARM, Lambda@Edge) and GCP for data retrieval (\cref{fig:cost_retrieval}), processing (\cref{fig:cost_processing}), and AI inference (\cref{fig:cost_inference}). GCP consistently offers lower costs, especially at scale, while AWS x86 and ARM have lower fixed costs but higher per-request expenses. While AWS Lambda@Edge is deployed closer to end-users and data sources, potentially reducing latency, in all three use cases, data retrieval (\cref{fig:cost_retrieval}), data processing (\cref{fig:cost_processing}), and AI inference (\cref{fig:cost_inference}), AWS Lambda@Edge incurs the highest per-request costs compared to AWS (x86 and ARM), and GCP. For AI inference, GCP's lower incremental costs make it more cost-efficient beyond the crossover point requests; despite its higher fixed cost, it remains more economical for sustained workloads due to its lower incremental cost.
\cref{fig:cost_cumulative} combines the cost of all functions to estimate the total workflow costs. Overall, if one considers the total workflow, AWS is cheaper for low-volume requests, and after crossing the point of around 9 million requests, GCP becomes more cost-efficient.

\cref{fig:cost_breakdown} shows the cost breakdown for AWS (x86, ARM, Lambda@Edge) and GCP across three workflows: data retrieval (\cref{fig:breakdown_a}), data processing (\cref{fig:breakdown_b}), and AI inference (\cref{fig:breakdown_c}). Costs are categorized into invocation, execution, backend-as-a-service (BaaS), data transfer, and state management.
For data retrieval, data transfer dominates, accounting for 53\% (AWS x86), 54\% (AWS ARM), and 75\% (Lambda@Edge), while GCP is at 52\%. 
For data processing, BaaS costs account for 44\% of AWS, and data transfer (37\% AWS; 12\% GCP) drives expenses. AWS Lambda@Edge remains the most expensive due to high execution costs.
For AI inference, BaaS dominates, comprising 83\% of AWS costs and 97\% of GCP.

\begin{tcolorbox}[colframe=gray, colback=gray!10, coltitle=black, boxrule=0.3mm,  rounded corners,left=1mm, right=1mm, top=1mm, bottom=1mm]
\textbf{\emph{Cost Provider Takeaway:}} GCP has 30\% lower costs for data-intensive and 57\% for compute-intensive workloads than AWS x86. AWS x86 has 75\% lower costs for AI inference at low workloads. AWS Lambda@Edge incurs 35\% higher data retrieval and 25\% higher data processing.
\end{tcolorbox}

\begin{tcolorbox}[colframe=gray, colback=gray!10, coltitle=black, boxrule=0.3mm,  rounded corners,left=1mm, right=1mm, top=1mm, bottom=1mm]
\textbf{\emph{Cost Driver Takeaway:}} In data retrieval functions, data transfer and state management drive costs, making up 53\% (AWS x86), 54\% (AWS ARM), and 75\% (AWS Lambda@Edge), compared to 52\% on GCP. In data processing, BaaS is the primary cost driver, accounting for 60\% of AWS x86 and ARM, and 48\% of AWS Lambda@Edge, while data transfer contributes 37\%, 38\%, and 29\%, respectively, versus 12\% on GCP. In AI inference, BaaS dominates, comprising 83\% of AWS costs and 97\% of GCP.
\end{tcolorbox}

\subsection{Threats to Validity}
Conducting experiments in space is still challenging due to the high cost, limited accessibility, and lack of publicly available pricing models for in-orbit processing. To the best of our knowledge, there is no standardized cost framework for LEO-based pay-per-use processing. Therefore, our evaluation focuses on edge-cloud-based serverless workflows, which provide a controlled environment that enables us to identify and characterize key cost drivers.
For LEO, we make assumptions about pricing models to be able to show the performance trade-off between latency and costs based on Takeme2Space\cite{tm2space_orbitlab}. While our findings provide a foundation for developing a framework that captures key cost-performance trade-offs, we acknowledge that they may vary as pricing models across the 3D Continuum evolve.

\section{Related Work}
\label{sec:rel_work}

Costless~\cite{Costless} analyzes the different factors that affect the pricing in serverless functions and presents a framework that predicts the cost and decides about function placement and fusion to minimize the costs. Costless cost model provides granular insights into AWS Lambda functions. However, their case study is tailored for one cloud provider and does not capture the dynamic characteristics of heterogeneous environments such as the 3D Continuum.
In~\cite{MLCosts2024}, authors present a cost model that estimates the total cost of geodistributed training in a multicloud environment, considering the compute cost, storage cost, and data transfer cost. Nevertheless, this cost model only considers storage as a BaaS service, neglecting additional services such as API gateways, costs specific to data transfers, and fixed costs associated with AI-related services. In ~\cite{PredictingCostsServerless2020}, the authors introduce a tool to predict and optimize the costs of serverless workflows without time-consuming experimentation. The proposed approach leverages Mixture Density Networks to model response time and Monte Carlo simulations to estimate the costs of entire workflows. Although this framework offers high accuracy, it considers the workflow as a whole, making it challenging to identify the most costly parts and how to minimize their impact on costs.
COSTA~\cite{COSTA} proposes an adaptive cost management framework that constantly monitors and migrates microservices applications across cloud providers to adjust to real-time pricing, reducing costs and avoiding cost SLO violations. However, such frameworks are tailored for VM applications, such as microservices, overlooking serverless-specific costs like BaaS and related services. It also focuses on runtime costs, neglecting fixed costs such as subscriptions or reserved instances, which can be significant for some functions. 

Unlike these frameworks that aggregate serverless workflow costs, Cosmos categorizes costs into fixed (e.g., monthly fees) and operational (e.g., invocation, compute, BaaS, storage, and data transfer), allowing precise cost breakdowns and targeted optimizations.

\section{Conclusion}
\label{sec:con}

In this paper, we introduce Cosmos, a cost model and performance-cost tradeoff model for serverless workflows, focusing on a detailed classification of main cost drivers such as invocation, compute, data transfer, state management, and BaaS. By analyzing serverless functions with different workload characteristics, such as data and compute intensity, our cost model provides a comprehensive understanding of serverless costs in the 3D Continuum. To validate our proposed cost model, we executed experiments on leading cloud platforms such as AWS and GCP.
The results show that data transfer and state management costs significantly account for 75\% of the costs of IO-intensive functions. On the other hand, BaaS costs dominate compute-intensive functions, accounting for as much as 97\%. While processing data closer to the function provides lower latency, it incurs up to 35\% higher costs.
These results highlight the need to align workload characteristics with the dynamic conditions of the 3D Continuum. A fine-grained cost classification is crucial for addressing workload-specific requirements while maximizing performance and minimizing operational costs. 
In future work, we plan to introduce different optimization mechanisms based on workload characteristics and compare them against state-of-the-art approaches. Additionally, we aim to develop an intelligent framework capable of predicting costs for individual functions in a workflow and dynamically selecting the most suitable layer and provider, accounting for workload-specific characteristics and SLOs, such as latency and cost.

\section*{Acknowledgment}
This work is partially funded by the Austrian Research Promotion Agency (FFG) under the project RapidREC (Project No. 903884).
This work has received funding from the Austrian Internet Stiftung under the NetIdee project LEO Trek (ID~7442).
This research received funding from the EU’s Horizon Europe Research and Innovation Program through TEADAL (GA No. 101070186) and NexaSphere projects (GA No. 101192912).

\balance

\bibliographystyle{IEEEtran}
\bibliography{references}

\end{document}